\documentclass[a4paper,twocolumn,english,prl,showpacs,superscriptaddress]{revtex4-1}

\usepackage[T1]{fontenc}
\usepackage[latin9]{inputenc}
\usepackage{fancyhdr}
\usepackage{babel}
\usepackage{textcomp}
\usepackage{amsmath}
\usepackage{amssymb}
\usepackage{graphicx}
\usepackage[unicode=true,pdfusetitle,
 bookmarks=true,bookmarksnumbered=false,bookmarksopen=false,
 breaklinks=false,pdfborder={0 0 1},backref=false,colorlinks=false]
 {hyperref}

\makeatletter

\pdfpageheight\paperheight
\pdfpagewidth\paperwidth

\providecommand{\tabularnewline}{\\}

\usepackage{babel}
\usepackage{calrsfs}
\usepackage{hyperref}
\hypersetup{
    colorlinks=true,
    linkcolor=red,
    citecolor=blue,
    filecolor=magenta,      
    urlcolor=cyan,
}
\makeatother

\begin{document}

\title{Unambiguous discrimination of nonorthogonal quantum states in cavity
QED}

\author{R. J. de Assis}

\affiliation{Instituto de Física, Universidade Federal de Goiás, 74.001-970, Goiânia
- GO, Brazil}

\author{J.S. Sales}

\affiliation{Centro de Ciências Exatas e Tecnológicas, Universidade Estadual de
Goiás, 75132-903, Anápolis, Goiás, Brazil}

\author{N. G. de Almeida}

\affiliation{Instituto de Física, Universidade Federal de Goiás, 74.001-970, Goiânia
- GO, Brazil}

\begin{abstract}
We propose an oversimplified scheme to unambiguously discriminate
nonorthogonal quantum field states inside high-Q cavities. Our scheme,
which is based on positive operator-valued measures (POVM) technique,
uses a single three-level atom interacting resonantly with a single
mode of a cavity-field and selective atomic state detectors. While
the single three-level atom takes the role of the ancilla, the single
cavity mode field represents the system\textbf{ }we want to obtain
information. The efficiency of our proposal is analyzed considering
the nowadays achievements in the context of cavity QED.
\end{abstract}
\maketitle

\section{Introduction}

\textit{\emph{Positive operator-valued measures (POVM) generalizes
all possible kind of measurements \cite{povm,jacobs14} and cannot
be reduced to standard projections of the initial state onto orthogonal
states spanning the initial Hilbert space alone, pertaining to the
system we want to obtain information }}\cite{Peres93,Nielsen00}\textit{\emph{.
In fact, although in general POVM can always be realized as standard
projective measurements on an enlarged system }}\cite{Peres93}\textit{\emph{,
they are such that the number of outputs may be larger than the dimensionality
of the space of states of the system in which we are interest in.
POVM is now standard in several areas of quantum mechanics, including
quantum optics and quantum information, among others \cite{Nielsen00,Scully97,Lo98,Janos07}.
In this paper we show how to accomplish POVM to unambiguously discriminate
nonorthogonal field states inside high-Q cavities. }}The goal of unambiguous
quantum state discrimination (UQSD) is to discern in which state the
system was prepared \cite{Ivanovic87,Chefles00,Cheffles12,Gisin96},
founding many applications in several protocols \cite{Chefles00,Cheffles12},
mainly for quantum cryptography \cite{Bennet17,Bennett84,Kak06,Eckert91}.\textbf{\textit{\emph{
}}}\textit{\emph{Our scheme, employing one three-level atom interacting
with a single mode of a cavity field, is very simple from the experimental
point of view and can be implemented using nowadays techniques in
cavity QED.}}

\textit{\emph{We begin by reviewing the general quantum measurement
theory. Next, we present our model and results, comparing with the
simple case of projective measurements. Then we present our conclusions.}}

\section{General measurements}

Consider a quantum system we are interested to measure, and a second
quantum system we call the \emph{ancilla},\emph{ }which is used to
get information about the system of interest \cite{jacobs14}. Let
the Hilbert space dimension of the system of interest and the ancilla
as $K$ and $L$, respectively. The ancilla is prepared in some known
initial state independently of the system of interest, and then the
two systems are allowed to interact, getting correlated.\textbf{ }Next,
a von Neumann measurement is performed on the ancilla, providing us
with information about the system of interest, which we are going
to call the\textbf{ }system from now on. Let us call the initial state
of the ancilla as $\left|a_{0}\right\rangle $ in the basis $\left\{ \left|a_{k}\right\rangle \right\} $,
$k=0,1,\ldots,K-1$, and denote the initial density operator of ancilla-system
as
\begin{equation}
\rho_{AS}=\left|a_{0}\right\rangle \left\langle a_{0}\right|\otimes\rho_{S},
\end{equation}
where $\rho_{S}$ is the initial density operator of the system. Let
$U$ denote the ancilla-system evolution operator. Since $U$ acts
in the tensor-product space, it can be written as
\begin{equation}
U=\sum_{kk'}\left|a_{k}\right\rangle \left\langle a_{k'}\right|\otimes M_{kk'}
\end{equation}
where 
\begin{equation}
M_{kk'}=\sum_{ll'}u_{klk'l'}\left|s_{l}\right\rangle \left\langle s_{l'}\right|,
\end{equation}
being $\left\{ \left|s_{l}\right\rangle \right\} $, $l=0,1,\ldots,L-1$,
a set of basis states for the system, and $u_{klk'l'}=\left\langle a_{k},s_{l}\right|U\left|a_{k'},s_{l'}\right\rangle $
are the matrix elements of $U$. Note that $M_{kk'}$ acts in the
Hilbert space of the system, and since the space of system has dimension
$L$, each sub-block matrix $M_{kk'}$ has dimension $L$. From now
on we use $M_{k}$ to refer to the first column of the sub-block $M_{k0}$
of $U$. Denoting the $K\times K$ sub-blocks of the matrix $U^{\dagger}U$
by $B_{kk'}$, and since $U^{\dagger}U=I$, it is readily seen that
\begin{equation}
B_{00}=\sum_{k}M_{k}^{\dagger}M_{k}=I.\label{eq:4}
\end{equation}
The important point to note here is that $M_{k}$ can be chosen to
be any set of operators, provided the restriction Eq. \eqref{eq:4}
above is obeyed. 

Now, performing a von Neumann measurement on the ancilla states, represented
by $\left|a_{m}\right\rangle \left\langle a_{m}\right|$, we can write
the (unnormalized) collapsed state of both ancilla and system as 
\begin{equation}
\widetilde{\rho}_{AS,m}=\left(\left|a_{m}\right\rangle \left\langle a_{m}\right|\otimes I\right)\rho_{AS}\left(\left|a_{m}\right\rangle \left\langle a_{m}\right|\otimes I\right)
\end{equation}
or, in terms of the sub-blocks of $U$:
\begin{equation}
\widetilde{\rho}_{AS,m}=\left|a_{m}\right\rangle \left\langle a_{m}\right|\otimes M_{m}\rho_{S}M_{m}^{\dagger}.\label{eq:6}
\end{equation}
From Eq. \eqref{eq:6} we can write the normalized state of the system
as 
\begin{equation}
\widetilde{\rho}_{S,m}=\frac{M_{m}\rho_{S}M_{m}^{\dagger}}{p_{m}},
\end{equation}
where 
\begin{equation}
p_{m}=\text{Tr}\left(M_{m}^{\dagger}M_{m}\rho_{S}\right)
\end{equation}
is the probability of finding the ancilla in state $\left|a_{m}\right\rangle $
after the unitary evolution $U$. It is now promptly recognized that
every set of operators $\left\{ M_{k}\right\} $ satisfying $\sum_{k}M_{k}^{\dagger}M_{k}\equiv\sum_{k}E_{k}=I$
describes a possible measurement on a quantum system, with the measuring
having $K$ outcomes. This gives us a complete description of a quantum
system under a general measurement. Next, we use the above results
to discriminate one of two nonorthogonal field states prepared into
a high Q cavity.

\section{Model}

\begin{figure}
\centering{}\includegraphics[bb=25bp 0bp 737bp 283bp,scale=0.36]{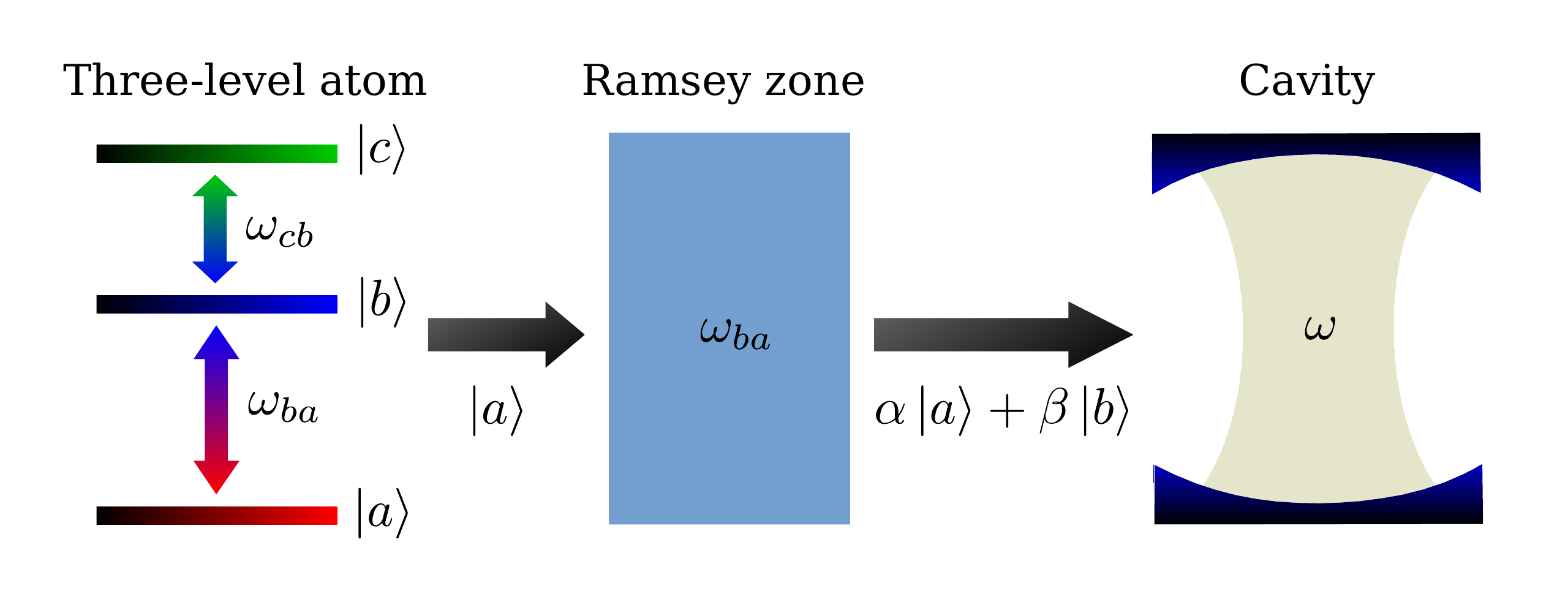}\caption{\label{fig:1}Scheme of the experimental setup to implement POVM in
cavity QED. A three-level atom interacts with a single mode of a high
Q cavity. The cavity mode field is prepared in one of two nonorthogonal
states $\left|\psi_{1}\right\rangle $ and $\left|\psi_{2}\right\rangle $.
Our POVM discriminates between these two nonorthogonal states prepared
inside the high Q cavity.}
\end{figure}

In our proposal, see Fig. \ref{fig:1}, a three-level atom in ladder
configuration, described by the set of states $\left\{ \left|a\right\rangle ,\left|b\right\rangle ,\left|c\right\rangle \right\} $,\textbf{
}is initially prepared in $\left|a\right\rangle $ and crosses a \textit{Ramsey
zone} (\textit{carrier interaction}).\textbf{ }Next, the atom enters
a cavity\textbf{ }interacting on-resonance with a singe mode of a
cavity field which in turns is prepared either in state $\left|\psi_{1}\right\rangle $
or in state $\left|\psi_{2}\right\rangle $, for which $\langle\psi_{1}\vert\psi_{2}\rangle\neq0$
(nonorthogonal states). 
While inside the cavity, the atom suffers a Stark shift in order to
lead $\omega_{ba}=\omega_{cb}=\omega$ \cite{Haroche99,Haroche01}.
After the atom crosses the cavity, it is detected in one of its three
possible states, thus revealing in which state the cavity mode was
prepared. The Hamiltonian model is given by \cite{Scully97} 

\begin{equation}
H=\hbar g_{1}\left(\sigma_{ba}a+\sigma_{ab}a^{\dagger}\right)+\hbar g_{2}\left(\sigma_{cb}a+\sigma_{bc}a^{\dagger}\right),\label{eq:9}
\end{equation}
where $\sigma_{\eta\nu}=\left|\eta\right\rangle \left\langle \nu\right|$
and $a^{\dagger}$ $\left(a\right)$ is the creation (annihilation)
photon number operator, and $g_{i}$ is the atom-field coupling, which
we take as real for convenience. In this protocol we are interested
in discriminating nonorthogonal states which are combinations of the
Fock states $\left|0\right\rangle $ and $\left|1\right\rangle $.
Thus, since the maximum number of photons in this case is $n=2$,
which happens when the atom decays and increases the photon number
into the cavity, we can consider $a^{\dagger}\left|2\right\rangle =0$.
After a little algebra, it is straightforward to obtain
\begin{widetext}
\begin{equation}
U\left|a,0\right\rangle =\alpha\left|a,0\right\rangle -i\beta\sin\left(g_{1}t\right)\left|a,1\right\rangle +\beta\cos\left(g_{1}t\right)\left|b,0\right\rangle ,\label{eq:10}
\end{equation}
\begin{multline}
U\left|a,1\right\rangle =\alpha\cos\left(g_{1}t_{c}\right)\left|a,1\right\rangle -i\beta\left(\frac{\sqrt{2}g_{1}}{\sqrt{2g_{1}^{2}+g_{2}^{2}}}\right)\sin\left(\sqrt{2g_{1}^{2}+g_{2}^{2}}t\right)\left|a,2\right\rangle -i\alpha\sin\left(g_{1}t\right)\left|b,0\right\rangle +\\
+\beta\cos\left(\sqrt{2g_{1}^{2}+g_{2}^{2}}t\right)\left|b,1\right\rangle -i\beta\left(\frac{g_{2}}{\sqrt{2g_{1}^{2}+g_{2}^{2}}}\right)\sin\left(\sqrt{2g_{1}^{2}+g_{2}^{2}}t\right)\left|c,0\right\rangle ,\label{eq:11}
\end{multline}
\begin{multline}
U\left|a,2\right\rangle =\alpha\left\{ 1+\left(\frac{2g_{1}^{2}}{2g_{1}^{2}+g_{2}^{2}}\right)\left[\cos\left(\sqrt{2g_{1}^{2}+g_{2}^{2}}t\right)-1\right]\right\} \left|a,2\right\rangle -i\alpha\left(\frac{\sqrt{2}g_{1}}{\sqrt{2g_{1}^{2}+g_{2}^{2}}}\right)\sin\left(\sqrt{2g_{1}^{2}+g_{2}^{2}}t\right)\left|b,1\right\rangle +\\
+\beta\cos\left(\sqrt{2}g_{2}t\right)\left|b,2\right\rangle +\alpha\left(\frac{\sqrt{2}g_{1}g_{2}}{2g_{1}^{2}+g_{2}^{2}}\right)\left[\cos\left(\sqrt{2g_{1}^{2}+g_{2}^{2}}t\right)-1\right]\left|c,0\right\rangle -i\beta\sin\left(\sqrt{2}g_{2}t\right)\left|c,1\right\rangle ,\label{eq:12}
\end{multline}
\end{widetext}

\noindent where $U=U_{C}U_{RZ}$, with $U_{C}=e^{-iHt/\hbar}$, $H$
given by Eq.\eqref{eq:9}, and $U_{ZR}$ is the evolution operator
as given by the \textit{\emph{carrier}} or \textit{\emph{Ramsey zone}}:
$U_{RZ}\left|a\right\rangle =\alpha\left|a\right\rangle +\beta\left|b\right\rangle $. 

Following the standard procedure \cite{Nielsen00}, we must build
three POVM elements: the one that discriminates $\left|\psi_{1}\right\rangle $;
the other one that discriminates $\left|\psi_{2}\right\rangle $,
and a third one leading to inconclusive results with probability $p_{in}$.
It is to be noted that the only constraint obeyed by the POVM elements
is $\sum_{k}E_{k}=I$. Thus, as soon as the atom state is known, we
will know \textit{\emph{with certainty}}\textit{ }that the cavity
mode field was either in state $\left|\psi_{1}\right\rangle $ or
in state $\left|\psi_{2}\right\rangle $, or that we do not know the
initial state as a result of the inconclusive measurement. As explained
above, to build the three POVM elements $E_{\nu}=M_{\nu}^{\dagger}M_{\nu}$,
$\nu=a,b,c$, we have to calculate $\left(n,n'=1,2,3\right)$:
\begin{eqnarray}
M_{\nu}= & \underset{nn'}{\sum}u_{\nu nan'}\left|n\right\rangle \left\langle n'\right| & ,\label{eq:13}
\end{eqnarray}
and
\begin{equation}
u_{\nu nan'}=\left\langle \nu,n\right|U\left|a,n'\right\rangle .
\end{equation}

Using Eq. \eqref{eq:10}-\eqref{eq:12}, we calculate the following
operators in Eq. \eqref{eq:13}:
\begin{widetext}
\begin{multline}
M_{a}=\alpha\left|0\right\rangle \left\langle 0\right|-i\beta\sin\left(g_{1}t\right)\left|1\right\rangle \left\langle 0\right|+\alpha\cos\left(g_{1}t\right)\left|1\right\rangle \left\langle 1\right|-\\
-i\beta\left(\frac{\sqrt{2}g_{1}}{\sqrt{2g_{1}^{2}+g_{2}^{2}}}\right)\sin\left(\sqrt{2g_{1}^{2}+g_{2}^{2}}t\right)\left|2\right\rangle \left\langle 1\right|+\alpha\left\{ 1+\left(\frac{2g_{1}^{2}}{2g_{1}^{2}+g_{2}^{2}}\right)\left[\cos\left(\sqrt{2g_{1}^{2}+g_{2}^{2}}t\right)-1\right]\right\} \left|2\right\rangle \left\langle 2\right|,\label{eq:15}
\end{multline}
\begin{multline}
M_{b}=\beta\cos\left(g_{1}t\right)\left|0\right\rangle \left\langle 0\right|-i\alpha\sin\left(g_{1}t\right)\left|0\right\rangle \left\langle 1\right|+\beta\cos\left(\sqrt{2g_{1}^{2}+g_{2}^{2}}t\right)\left|1\right\rangle \left\langle 1\right|-\\
-i\alpha\left(\frac{\sqrt{2}g_{1}}{\sqrt{2g_{1}^{2}+g_{2}^{2}}}\right)\sin\left(\sqrt{2g_{1}^{2}+g_{2}^{2}}t\right)\left|1\right\rangle \left\langle 2\right|+\beta\cos\left(\sqrt{2}g_{2}t\right)\left|2\right\rangle \left\langle 2\right|,\label{eq:16}
\end{multline}
\begin{multline}
M_{c}=-i\beta\left(\frac{g_{2}}{\sqrt{2g_{1}^{2}+g_{2}^{2}}}\right)\sin\left(\sqrt{2g_{1}^{2}+g_{2}^{2}}t\right)\left|0\right\rangle \left\langle 1\right|+\\
+\alpha\left(\frac{\sqrt{2}g_{1}g_{2}}{2g_{1}^{2}+g_{2}^{2}}\right)\left[\cos\left(\sqrt{2g_{1}^{2}+g_{2}^{2}}t\right)-1\right]\left|0\right\rangle \left\langle 2\right|-i\beta\sin\left(\sqrt{2}g_{2}t\right)\left|1\right\rangle \left\langle 2\right|.\label{eq:17}
\end{multline}
\end{widetext}

\noindent From Eq. \eqref{eq:15}-\eqref{eq:17} we can calculate
the POVM elements $E_{\nu}=M_{\nu}^{\dagger}M_{\nu}$ for $\nu=a,b,c$:
\begin{widetext}
\begin{multline}
E_{a}=\left[\left|\alpha\right|^{2}+\left|\beta\right|^{2}\sin^{2}\left(g_{1}t\right)\right]\left|0\right\rangle \left\langle 0\right|+i\sin\left(g_{1}t\right)\cos\left(g_{1}t\right)\left(\alpha\beta^{*}\left|0\right\rangle \left\langle 1\right|-\alpha^{*}\beta\left|1\right\rangle \left\langle 0\right|\right)+\\
+\left[\left|\alpha\right|^{2}\cos^{2}\left(g_{1}t\right)+\left|\beta\right|^{2}\left(\frac{2g_{1}^{2}}{2g_{1}^{2}+g_{2}^{2}}\right)\sin^{2}\left(\sqrt{2g_{1}^{2}+g_{2}^{2}}t\right)\right]\left|1\right\rangle \left\langle 1\right|+\\
+i\left(\frac{\sqrt{2}g_{1}}{\sqrt{2g_{1}^{2}+g_{2}^{2}}}\right)\left\{ 1+\left(\frac{2g_{1}^{2}}{2g_{1}^{2}+g_{2}^{2}}\right)\left[\cos\left(\sqrt{2g_{1}^{2}+g_{2}^{2}}t\right)-1\right]\right\} \left(\alpha\beta^{*}\left|1\right\rangle \left\langle 2\right|-\alpha\beta^{*}\left|2\right\rangle \left\langle 1\right|\right)+\\
+\left|\alpha\right|^{2}\left\{ 1+\left(\frac{2g_{1}^{2}}{2g_{1}^{2}+g_{2}^{2}}\right)\left[\cos\left(\sqrt{2g_{1}^{2}+g_{2}^{2}}t\right)-1\right]\right\} ^{2}\left|2\right\rangle \left\langle 2\right|,
\end{multline}
\begin{multline}
E_{b}=\left|\beta\right|^{2}\cos^{2}\left(g_{1}t\right)\left|0\right\rangle \left\langle 0\right|-i\sin\left(g_{1}t\right)\cos\left(g_{1}t\right)\left(\alpha\beta^{*}\left|0\right\rangle \left\langle 1\right|-\alpha\beta^{*}\left|1\right\rangle \left\langle 0\right|\right)+\\
\\
+\left[\left|\alpha\right|^{2}\sin^{2}\left(g_{1}t\right)+\left|\beta\right|^{2}\cos^{2}\left(\sqrt{2g_{1}^{2}+g_{2}^{2}}t\right)\right]\left|1\right\rangle \left\langle 1\right|-\\
-i\left(\frac{\sqrt{2}g_{1}}{\sqrt{2g_{1}^{2}+g_{2}^{2}}}\right)\sin\left(\sqrt{2g_{1}^{2}+g_{2}^{2}}t\right)\cos\left(\sqrt{2g_{1}^{2}+g_{2}^{2}}t\right)\left(\alpha\beta^{*}\left|1\right\rangle \left\langle 2\right|-\alpha\beta^{*}\left|2\right\rangle \left\langle 1\right|\right)+\\
+\left[\left|\alpha\right|^{2}\left(\frac{2g_{1}^{2}}{2g_{1}^{2}+g_{2}^{2}}\right)\sin^{2}\left(\sqrt{2g_{1}^{2}+g_{2}^{2}}t\right)+\left|\beta\right|^{2}\cos^{2}\left(\sqrt{2}g_{2}t\right)\right]\left|2\right\rangle \left\langle 2\right|,
\end{multline}
\begin{multline}
E_{c}=\left|\beta\right|^{2}\left(\frac{g_{2}^{2}}{2g_{1}^{2}+g_{2}^{2}}\right)\sin^{2}\left(\sqrt{2g_{1}^{2}+g_{2}^{2}}t\right)\left|1\right\rangle \left\langle 1\right|+\\
+i\left[\frac{\sqrt{2}g_{1}g_{2}^{2}}{\left(2g_{1}^{2}+g_{2}^{2}\right)^{\frac{3}{2}}}\right]\sin\left(\sqrt{2g_{1}^{2}+g_{2}^{2}}t\right)\left[\cos\left(\sqrt{2g_{1}^{2}+g_{2}^{2}}t\right)-1\right]\left(\alpha\beta^{*}\left|1\right\rangle \left\langle 2\right|-\alpha\beta^{*}\left|2\right\rangle \left\langle 1\right|\right)+\\
+\left\{ \left|\alpha\right|^{2}\left[\frac{2g_{1}^{2}g_{2}^{2}}{\left(2g_{1}^{2}+g_{2}^{2}\right)^{2}}\right]\left[\cos\left(\sqrt{2g_{1}^{2}+g_{2}^{2}}t\right)-1\right]^{2}+\left|\beta\right|^{2}\sin^{2}\left(\sqrt{2}g_{2}t\right)\right\} \left|2\right\rangle \left\langle 2\right|.
\end{multline}
\end{widetext}

\noindent As can be checked, $\sum_{\nu}E_{\nu}=I$.

To be specific, let us assume that we want to discriminate the following
nonorthogonal field states into the cavity: $\left|\psi_{1}\right\rangle =\left|0\right\rangle $
and $\left|\psi_{2}\right\rangle =\frac{1}{\sqrt{2}}\left(\left|0\right\rangle +\left|1\right\rangle \right)$
\cite{Nielsen00}. The cavity state is thus represented by $\rho=q_{1}\left|\psi_{1}\right\rangle \left\langle \psi_{1}\right|+q_{2}\left|\psi_{2}\right\rangle \left\langle \psi_{2}\right|$,
where $q_{1}$ $\left(q_{2}\right)$ is the classical probability
related to the frequency of preparing the state $\left|\psi_{1}\right\rangle $
$\left(\left|\psi_{2}\right\rangle \right)$ and $q_{1}+q_{2}=1$.
Clearly, $E_{c}$ discriminates state $\left|\psi_{2}\right\rangle $,
since $\left\langle \psi_{1}\right|E_{c}\left|\psi_{1}\right\rangle =0$.
To discriminate $\left|\psi_{1}\right\rangle $, we impose $\left\langle \psi_{2}\right|E_{b}\left|\psi_{2}\right\rangle =0$.
This imposition leads us with the conditions: (i) $\alpha=\cos\left(g_{1}t\right)$,
(ii) $\beta=i\sin\left(g_{1}t\right)$, and (iii) $\cos\left(\sqrt{2g_{1}^{2}+g_{2}^{2}}t\right)=0$.
Letting $g_{2}=\kappa g_{1}$, the third condition can be rewritten
as $g_{1}t\equiv\theta_{m}=\frac{(m+\frac{1}{2})\pi}{\sqrt{2+\kappa^{2}}}$,
$m=0,1,2,...$. On the other hand, $E_{a}$ is inconclusive, since
$\left\langle \psi_{1}\right|E_{a}\left|\psi_{1}\right\rangle \neq0$
and $\left\langle \psi_{2}\right|E_{a}\left|\psi_{2}\right\rangle \neq0$,
meaning that we must discard this measurement. Using these three conditions,
we can write the effective POVM elements in the following way:
\begin{multline}
E_{a}=\left(\cos^{2}\theta_{m}+\sin^{4}\theta_{m}\right)\left|0\right\rangle \left\langle 0\right|\\
+\frac{1}{4}\sin^{2}\left(2\theta_{m}\right)\left(\left|0\right\rangle \left\langle 1\right|+\left|1\right\rangle \left\langle 0\right|\right)+\\
+\left[\cos^{4}\theta_{m}+\left(\frac{2}{2+\kappa^{2}}\right)\sin^{2}\theta_{m}\right]\left|1\right\rangle \left\langle 1\right|,
\end{multline}

\begin{align}
E_{b} & =\frac{1}{2}\sin^{2}\left(2\theta_{m}\right)\left|\psi_{2}^{\perp}\right\rangle \left\langle \psi_{2}^{\perp}\right|,
\end{align}
\begin{align}
E_{c} & =\left(\frac{\kappa^{2}}{2+\kappa^{2}}\right)\sin^{2}\theta_{m}\left|1\right\rangle \left\langle 1\right|,
\end{align}
where we have neglected terms containing the state $\left|2\right\rangle $
and put $\left|\psi_{2}^{\perp}\right\rangle =\frac{1}{\sqrt{2}}\left(\left|0\right\rangle -\left|1\right\rangle \right)$. 

The probabilities related to the success probability rates of POVM
elements $E_{b}$ and $E_{c}$ are, respectively,
\begin{equation}
p_{b}=\text{Tr}\left(E_{b}\rho\right)=\frac{q_{1}}{4}\sin^{2}\left(2\theta_{m}\right)\label{eq:24}
\end{equation}
and

\begin{equation}
p_{c}=\text{Tr}\left(E_{c}\rho\right)=\frac{q_{2}}{2}\left(\frac{\kappa^{2}}{2+\kappa^{2}}\right)\sin{}^{2}\left(\theta_{m}\right),\label{eq:25}
\end{equation}
while the probability for the inconclusive results $p_{in}=p_{a}=\text{Tr}\left(E_{a}\rho\right)=q_{1}\left\langle \psi_{1}\right|E_{a}\left|\psi_{1}\right\rangle +q_{2}\left\langle \psi_{2}\right|E_{a}\left|\psi_{2}\right\rangle $
is \textbf{
\begin{multline}
p_{in}=q_{1}\left(\cos^{2}\theta_{m}+\sin^{4}\theta_{m}\right)+\\
+\frac{q_{2}}{2}\left[1+\cos^{2}\theta_{m}+\left(\frac{2}{2+\kappa^{2}}\right)\sin^{2}\theta_{m}\right].\label{eq:26}
\end{multline}
}The success probability is given by $p_{s}=p_{b}+p_{c}$:
\begin{equation}
p_{s}=\frac{q_{1}}{4}\sin^{2}\left(2\theta_{m}\right)+\frac{q_{2}}{2}\left(\frac{\kappa^{2}}{2+\kappa^{2}}\right)\sin{}^{2}\left(\theta_{m}\right),\label{eq:27}
\end{equation}
where $p_{s}+p_{in}=1$, as should. 

\section{Discussion}

\begin{figure}[ptb]
\centering{}\includegraphics[bb=15bp 0bp 576bp 432bp,scale=0.47]{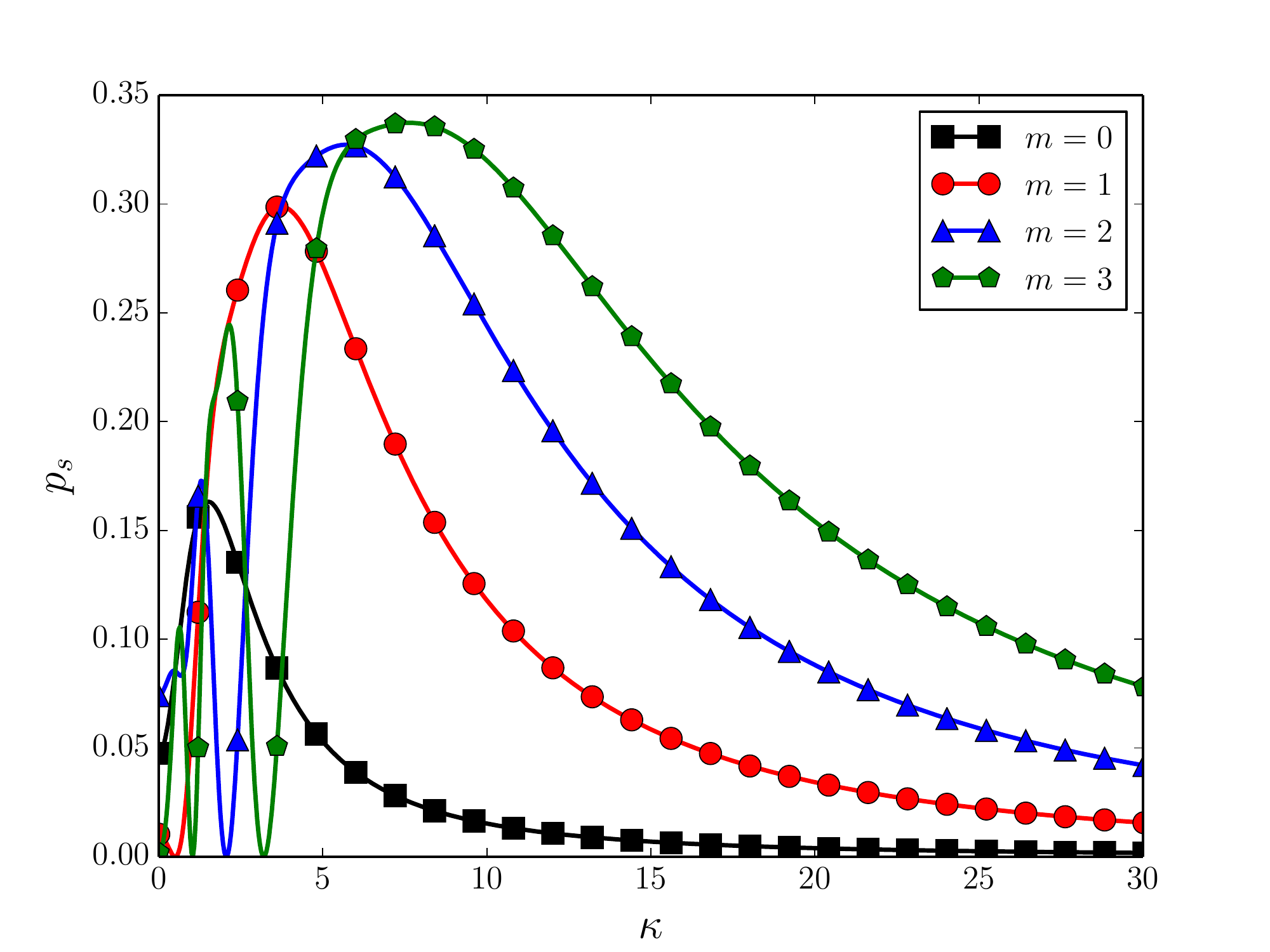}\caption{\label{fig:2}Success probability $p_{s}$ \emph{versus} the coupling
ratio $\kappa=g_{2}/g_{1}$ for $q_{1}=0.3$. The best choice to UQED
is the one that maximizes $p_{s}$ to each integer $m$. }
\end{figure}

\begin{figure}[ptb]
\centering{}\includegraphics[bb=15bp 0bp 576bp 432bp,scale=0.47]{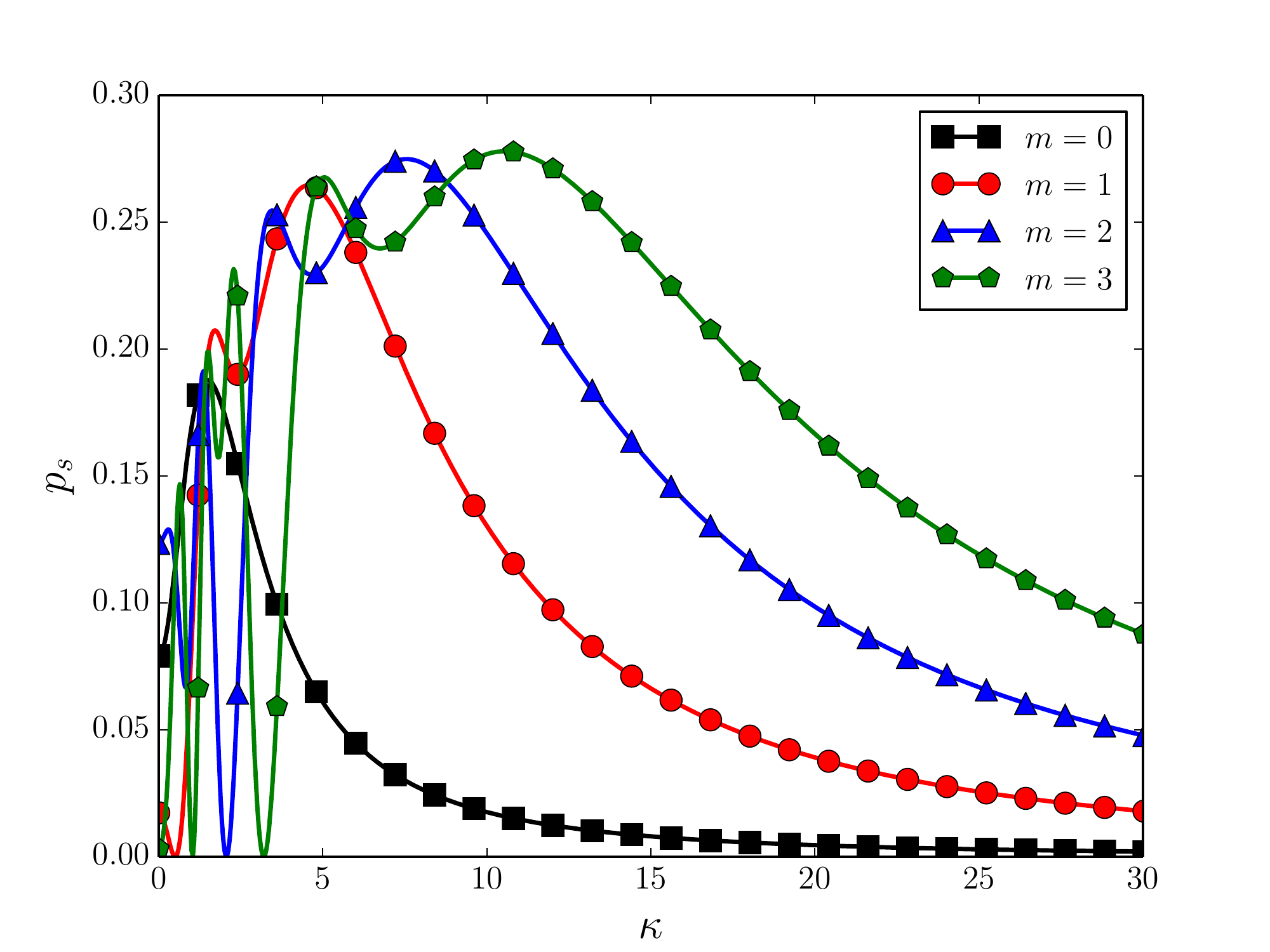}\caption{\label{fig:3}Success probability $p_{s}$ \emph{versus} the coupling
ratio $\kappa=g_{2}/g_{1}$ for $q_{1}=0.7$. The best choice to UQED
is the one that maximizes $p_{s}$ to each integer $m$.}
\end{figure}

\begin{figure}[ptb]
\centering{}\includegraphics[bb=15bp 0bp 576bp 432bp,scale=0.47]{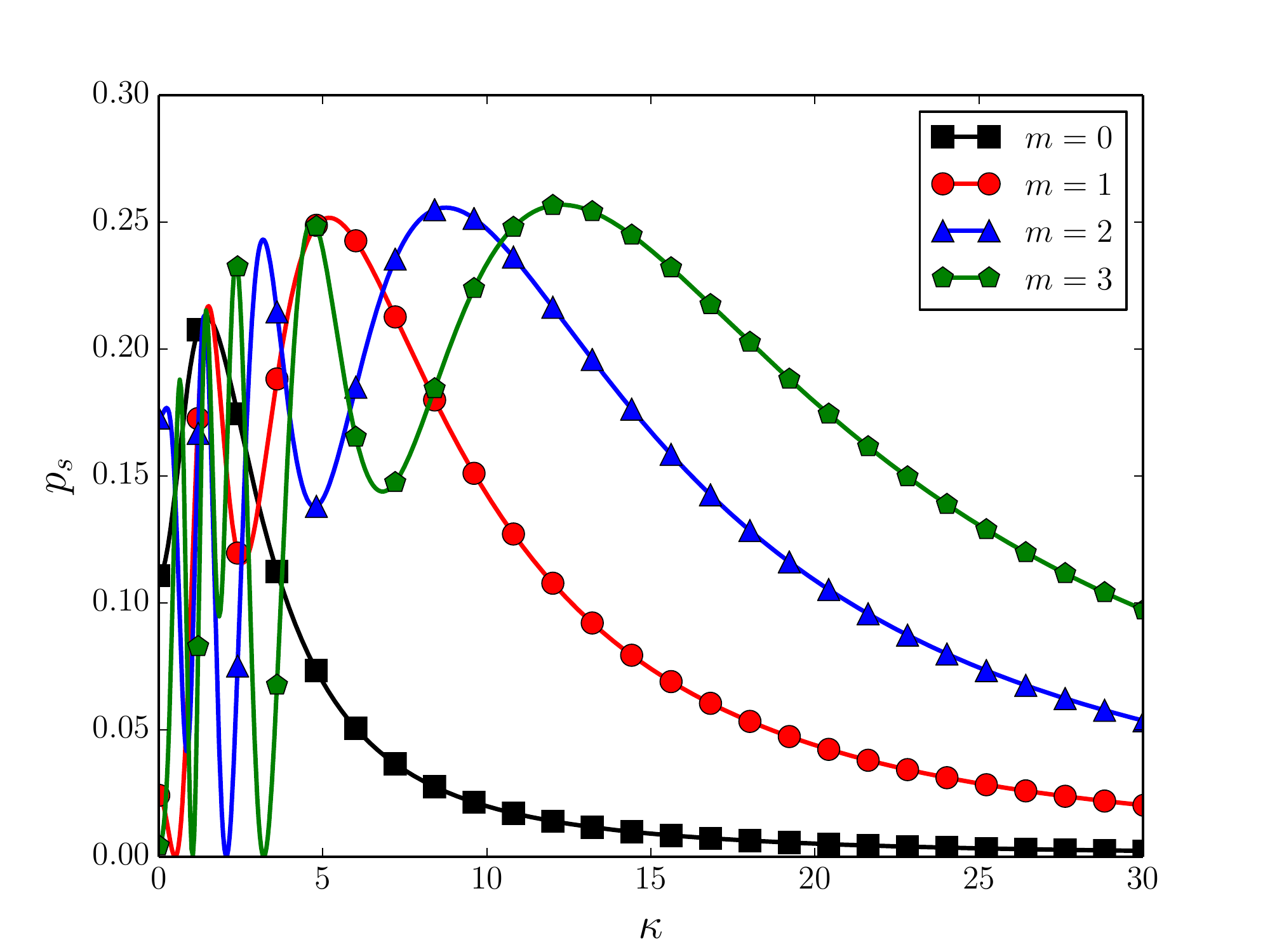}\caption{\label{fig:4}Success probability $p_{s}$ \emph{versus} the coupling
ratio $\kappa=g_{2}/g_{1}$ for $q_{1}=0.5$. The best choice to UQED
is the one that maximizes $p_{s}$ to each integer $m$. }
\end{figure}

Since the result of POVM element $E_{a}$ is the inconclusive one,
all we have to do in order to optimize our proposal is either minimize
$p_{in}$ or maximize $p_{s}$. We numerically maximize Eq. \eqref{eq:27}
with (i) $q_{1}=0.3$, (ii) $q_{1}=0.7$ and, for comparison to other
work, (iii) $q_{1}=q_{2}=0.5$ \cite{Nielsen00}. As an example, to
$q_{1}=q_{2}=0.5$ we find, see Fig. \ref{fig:4}, (a) for $m=0,$
$p_{s}=0.1878$ (black line with squares), implying $\kappa=1,47$,
(b) for $m=1$, $p_{s}=0.2644$ (red line with circles), implying
$\kappa=4.50$ (c) for $m=2,$ \textbf{$p_{s}=0.2748$ }(blue line
with triangles), implying $\kappa=7,50$ (d) for $m=3$, $p_{s}=0.2779$
(green line pentagons), implying $\kappa=10.55$. Values of $m\geq3$
can be used at the expense of greater ratio $\kappa=g_{2}/g_{1}$,
see Tab. \ref{tab:1}-\ref{tab:3}. The best choice to UQSD is the
one that minimizes (maximizes) $p_{in}$ $\left(p_{s}\right)$ for
each integer $m$, and it is worthwhile to mention that the success
probability rate around 0.26, obtained for the first values of the
integer m, is very close to the best rate of success for this kind
of quantum state discrimination predicted theoretically, which is
0.292 when $q_{1}=q_{2}$ \cite{Nielsen00,Janos07}. As can be seen
from Fig. \ref{fig:2}-\ref{fig:3}, there are several maxima in $p_{s}$,
depending on the value of $\kappa$. Here we have chosen those whose
success probability is the greatest one. Note from Tab. \ref{tab:3},
corresponding to $q_{1}=q_{2}=0.5$, that the best value for the success
probability rate is below 0.292, which is the maximum value according
to Ref. \cite{Nielsen00,Janos07}. This is also confirmed by our numerical
calculations using much greater values for $m$ and $\kappa$. 

It is to be noted that the simple strategy of choosing whether to
project the cavity field state on the computational basis $\left|0\right\rangle $
or $\left|1\right\rangle $ would allow us to discriminate only one
state. Indeed, if the result of projection is $\left|1\right\rangle $,
the cavity mode state could not have been prepared in $\left|\psi_{1}\right\rangle $
and was prepared therefore in state $\left|\psi_{2}\right\rangle $;
however, if the measurement result is $\left|0\right\rangle $, we
can not be sure if the cavity mode state had been prepared in $\left|\psi_{1}\right\rangle $
or $\left|\psi_{2}\right\rangle $, this result being inconclusive.
As a result of this strategy, we would find a success probability
of $p_{s}=\text{Tr}\left(\left|1\right\rangle \left\langle 1\right|\rho\right)=0.25$,
thus lesser than the POVM strategy we developed. In Tab. \ref{tab:1}-\ref{tab:3}
we present the values of $\kappa$ maximizing $p_{s}$ and the corresponding
values for $p_{in}$, $p_{b}$, $p_{c}$ and $p_{s}$, Eq. \eqref{eq:24}-\eqref{eq:27},
for several integers $m$ and $q_{1}=0.3,0.7,0.5$. 
\begin{table}
\begin{centering}
\begin{tabular}{|c|c|c|c|c|c|}
\hline 
$m$ & $\kappa$ & $p_{in}$ & $p_{b}$ & $p_{c}$ & $p_{s}$\tabularnewline
\hline 
\hline 
0 & 1.52 & 0.8369 & 0.0747 & 0.0884 & 0.1631\tabularnewline
\hline 
1 & 3.72 & 0.7011 & 0.0366 & 0.2623 & 0.2989\tabularnewline
\hline 
2 & 5.71 & 0.6728 & 0.0155 & 0.3117 & 0.3272\tabularnewline
\hline 
3 & 7.63 & 0.6627 & 0.0069 & 0.3304 & 0.3373\tabularnewline
\hline 
4 & 9.54 & 0.6581 & 0.0032 & 0.3387 & 0.3419\tabularnewline
\hline 
5 & 11.44 & 0.6556 & 0.0015 & 0.3429 & 0.3444\tabularnewline
\hline 
10 & 21.23 & 0.6505 & 0.0012 & 0.3483 & 0.3495\tabularnewline
\hline 
20 & 41.11 & 0.6505 & 0.0001 & 0.3494 & 0.3495\tabularnewline
\hline 
50 & 101.05 & 0.6501 & 0.0000 & 0.3499 & 0.3499\tabularnewline
\hline 
\end{tabular}
\par\end{centering}
\caption{\label{tab:1}Values for the probabilities according to our protocol
to accomplish UQSD in cavity QED. To each integer $m$, there is a
minimum for the inconclusive events $p_{in}$ and a corresponding
maximum for the probability of success $p_{s}$. The table shows $p_{in}$,
$p_{b}$, $p_{c}$, and $p_{s}$ separately separately for $p_{1}=0.3$.}
\end{table}
\begin{table}
\begin{centering}
\begin{tabular}{|c|c|c|c|c|c|}
\hline 
$m$ & $\kappa$ & $p_{in}$ & $p_{b}$ & $p_{c}$ & $p_{s}$\tabularnewline
\hline 
\hline 
0 & 1.45 & 0.7874 & 0.1749 & 0.0377 & 0.2126\tabularnewline
\hline 
1 & 5.19 & 0.7483 & 0.1693 & 0.0824 & 0.2517\tabularnewline
\hline 
2 & 8.54 & 0.7443 & 0.1679 & 0.0878 & 0.2557\tabularnewline
\hline 
3 & 12.27 & 0.7432 & 0.1674 & 0.0894 & 0.2568\tabularnewline
\hline 
4 & 15.80 & 0.7427 & 0.1673 & 0.0900 & 0.2573\tabularnewline
\hline 
5 & 19.32 & 0.7426 & 0.1671 & 0.0903 & 0.2574\tabularnewline
\hline 
10 & 36.91 & 0.7422 & 0.1670 & 0.0908 & 0.2578\tabularnewline
\hline 
20 & 72.08 & 0.7420 & 0.1670 & 0.0910 & 0.2580\tabularnewline
\hline 
50 & 177.58 & 0.7420 & 0.1670 & 0.0910 & 0.2580\tabularnewline
\hline 
\end{tabular}
\par\end{centering}
\caption{\label{tab:2}Values for the probabilities according to our protocol
to accomplish UQSD in cavity QED. To each integer $m$, there is a
minimum for the inconclusive events $p_{in}$ and a corresponding
maximum for the probability of success $p_{s}$. The table shows $p_{in}$,
$p_{b}$, $p_{c}$, and $p_{b}$ separately separately for $p_{1}=0.7$.}
\end{table}
\begin{table}
\begin{centering}
\begin{tabular}{|c|c|c|c|c|c|}
\hline 
$m$ & $\kappa$ & $p_{in}$ & $p_{b}$ & $p_{c}$ & $p_{s}$\tabularnewline
\hline 
\hline 
0 & 1.47 & 0.8123 & 0.1248 & 0.0629 & 0.1877\tabularnewline
\hline 
1 & 4.50 & 0.7356 & 0.1039 & 0.1605 & 0.2644\tabularnewline
\hline 
2 & 7.55 & 0.7252 & 0.0989 & 0.1759 & 0.2748\tabularnewline
\hline 
3 & 10.55 & 0.7221 & 0.0957 & 0.1822 & 0.2779\tabularnewline
\hline 
4 & 13.70 & 0.7209 & 0.0956 & 0.1835 & 0.2791\tabularnewline
\hline 
5 & 16.50 & 0.7101 & 0.0950 & 0.1849 & 0.2799\tabularnewline
\hline 
10 & 31.52 & 0.7191 & 0.0941 & 0.1868 & 0.2809\tabularnewline
\hline 
20 & 61.50 & 0.7189 & 0.0938 & 0.1873 & 0.2811\tabularnewline
\hline 
50 & 151.50 & 0.7188 & 0.0930 & 0.1882 & 0.2812\tabularnewline
\hline 
\end{tabular}
\par\end{centering}
\caption{\label{tab:3}Values for the probabilities according to our protocol
to accomplish UQSD in cavity QED. To each integer $m$, there is a
minimum for the inconclusive events $p_{in}$ and a corresponding
maximum for the probability of success $p_{s}$. The table shows $p_{in}$,
$p_{b}$, $p_{c}$, and $p_{s}$ separately separately for $p_{1}=0.5$.}
\end{table}

As a final remark, one could ask why to use POVM strategy instead
of simply projecting on the computational basis of the cavity states.
Three remarks are in order: (i) First, in addition to our protocol
presenting a higher probability of success, there is no known technique
to directly project the cavity state on the computational basis: usually,
measurement of the cavity state requires additional atoms and/or cavities,
thus being necessary to build another POVM elements to measure the
cavity mode field \cite{Iara07,Almeida11}, (ii) second, the direct
projective strategy does not discriminate both states but just $\left|\psi_{1}\right\rangle =\left|0\right\rangle $,
while the POVM strategy allows us to discriminate both $\left|\psi_{1}\right\rangle $
and $\left|\psi_{2}\right\rangle $, (iii) the POVM strategy was build
using known matter-radiation interaction parameters: it remains an
open question if other types of interaction, such as those developed
by effective Hamiltonian techniques \cite{Serra05,Prado06,deAlmeida14},
could attain optimal POVM results \cite{Jaeger95}.

\section{Conclusion}

We have proposed an oversimplified scheme to build POVM elements allowing
to discriminate nonorthogonal field states inside a high Q cavity.
Besides to circumvent the impossibility to directly project the cavity
states onto the computational Fock states without using ancilla, our
protocol achieves a rate of success probability greater than the direct
projective technique. Our proposal relies on nowadays techniques in
the cavity QED domain, making use of just one single three-level atom
undergoing a Ramsey zone (carrier interaction) plus one cavity and
selective atomic state detectors. This simplicity makes our protocol
very attractive from the experimental point of view. Finally, we hope
our protocol can inspire other POVM strategies based on effective
Hamiltonians technique making possible to attain optimal rates of
success probability. 
\begin{acknowledgments}
We acknowledge financial support from the Brazilian agency CNPq, CAPES
and FAPEG. This work was performed as part of the Brazilian National
Institute of Science and Technology (INCT) for Quantum Information.
\end{acknowledgments}


\begin{thebibliography}{10}
\bibitem{povm}C. W. Helstrom, Quantum Detection and Estimation Theory
(Academic, New York, 1976).

\bibitem{jacobs14}K. Jacobs, Quantum Measurement Theory and its Applications
(Cambridge University Press, 2014).

\bibitem{Peres93}A. Peres, Quantum Theory: Concepts and Methods (Kluwer,
Dordrecht, 1993), Chap. 9.

\bibitem{Nielsen00}Nielsen, M.A., Chuang, I.L.: Quantum Computation
and Quantum Information. Cambridge University Press, Cambridge (2000).

\bibitem{Scully97}M. O. Scully and M. S. Zubairy Quantum Optics (Cambridge
1997).

\bibitem{Lo98}H.-K. Lo, S. Popescu and T. P. Spiller (eds.), Introduction
to Quantum Computation and Information, (World Scientific Publishing,
1998), ISBN 981-02-3399-X.

\bibitem{Janos07}Bergou J. A., Quantum state discrimination and selected
applications. Journal of Physics: Conference series 84 012001 (2007).

\bibitem{Ivanovic87}I. D. Ivanovic, Physics Letters A Volume 123,
Issue 6, Pages 257-312 (17 August 1987).

\bibitem{Chefles00}A. Chefles, Quantum State Discrimination, Contemp.
Phys. 41, 401 (2000).

\bibitem{Cheffles12}Unambiguous discrimination among oracle operators
Anthony Chefles, Akira Kitagawa, Masahiro Takeoka, Masahide Sasaki,
and Jason Twamley, Journal of Physics A: Mathematical and Theoretical,
Volume 40, Issue 33, pp. 10183-10213 (2007).

\bibitem{Gisin96}B. Huttner, A. Muller, J. D. Gautier, H. Zbinden,
and N. Gisin, Phys. Rev. a 54, 5 1996).

\bibitem{Bennet17}C. H. Bennett, and G. Brassard \textquotedbl{}Quantum
cryptography: Public key distribution and coin tossing\textquotedbl{}.
Theoretical Computer Science, Vol. 560, Part 1, 4 December 2014, Pages
7\textendash 11.

\bibitem{Bennett84}Bennett, C.H. and G. Brassard. Quantum cryptography:
Public key distribution and coin tossing. In Proceedings of IEEE International
Conference on Computers, Systems and Signal Processing, volume 175,
page 8. New York, 1984.

\bibitem{Kak06}S. Kak, A three-stage quantum cryptography protocol.
Foundations of Physics Letters, vol. 19, pp. 293\textendash 296, 2006.

\bibitem{Eckert91}Ekert. A. Physical Review Letters, 67, pp. 661\textendash 663,
(1991).

\bibitem{Haroche99}S. Haroche, Conf. Proc. 464, 45 (1999).

\bibitem{Haroche01}A. Rauschenbeutel et al., Phys. Rev. A , 64 050301
(2001).

\bibitem{Iara07} Iara P de Queirós ; Simone Souza, W. B. Cardoso,
and N. G. de Almeida, Physical Review. A, v. 76, p. 0341011-0341014,
2007.

\bibitem{Almeida11} N. G. de Almeida, M. H. Y. Moussa, and R d J
Napolitano, J. of Phys. B, Atomic Molecular and Optical Physics (Print),
v. 44, p. 165502, 2011.

\bibitem{Serra05}R. M. Serra, C. J. Villas-Boas, ; N. G. de Almeida,
M. H. Y. Moussa, Frequency up- and down-conversion in two mode cavity
quantum electrodynamics. Phys. Rev. A 71, 45802 (2005). 

\bibitem{Prado06}F. O. Prado, N. G. de Almeida, M. H. Y. Moussa,
C. J. Villas-Boas, Bilinear and quadratic Hamiltonians in two-mode
cavity quantum electrodynamics. Phys. Rev.. A 73, 43803 (2006).

\bibitem{deAlmeida14}N. G. de Almeida, Engineering the unitary charge
conjugation operator of quantum field theory for particle-antiparticle
using trapped ions and light fields in cavity QED. Journal of Physics.
B, Atomic Molecular and Optical Physics, v. 47, p. 165501, 2014.

\bibitem{Jaeger95}G. Jaeger and A. Shimony: Phys. Lett. A 197, 83
(1995).
\end{thebibliography}
\end{document}